\documentclass[pra,aps,twocolumn,showpacs,amsmath,amssymb]{revtex4}
\usepackage{graphicx}
\usepackage{dcolumn}
\usepackage{bm}

\newcommand{\nn}{\nonumber}

\begin{document}

\title{Magnetization profile and core level spectroscopy in a multiply quantized vortex of imbalanced Fermi superfluids}

\author{K. M. Suzuki}
\email{kenta@mp.okayama-u.ac.jp}
\affiliation{Department of Physics, Okayama University,
Okayama 700-8530, Japan}
\author{T. Mizushima}
\email{mizushima@mp.okayama-u.ac.jp}
\affiliation{Department of Physics, Okayama University,
Okayama 700-8530, Japan}
\author{M. Ichioka}
\affiliation{Department of Physics, Okayama University,
Okayama 700-8530, Japan}
\author{K. Machida}
\affiliation{Department of Physics, Okayama University,
Okayama 700-8530, Japan}
\date{\today}

\begin{abstract}
The core structure of multiply quantized vortices is theoretically investigated in fermionic superfluid near Feshbach resonance. Under population imbalance in two hyperfine spin states, the vortex core is filled in by the ``paramagnetic moment''. Here, we find the spatial oscillation of the magnetization inside the core sensitively due to the topological structure of the pairing field, in the range from the weak coupling regime to the unitary limit. This magnetization inside the giant core reveals the winding number of the vortex and directly results from the low-lying quasiparticle states bound inside the core. It is therefore proposed that the density profile experiment using phase contrast imaging can provide the spectroscopy of novel core level structures in giant vortices. To help the understanding on these outcomes, we also derive the analytic solution for the low-lying quasiparticle states inside the core of a multiply quantized vortex.
\end{abstract}


\maketitle

\section{Introduction}

Much attention has been focused on quantized vortices of both bosonic and fermionic superfluids. In the bosonic system, the order parameter is directly associated with the particle density, which is observable via the absorption image in experiments. Hence, the vortex core, defined as the zeros of the order parameter, can be visualized via the absorption image in the Bose-Einstein condensate (BEC) of ultracold atoms, which has provided an opportunity to investigate both the static and dynamic properties of vortices \cite{fetter, kasamatsu}. 

In the fermionic case, however, the situation is drastically changed, where the order parameter corresponds to the wave function of the Cooper pair and thus the visualization of vortices is not trivial. The particle density is sensitively affected by the quasiparticle structure with the eigenenergy close to the Fermi energy $E_F$. The pioneering work in 1964 \cite{cdgm} revealed the fact that for a singly quantized vortex state in an $s$-wave case, there exists the novel quasiparticle state tightly bound in the core region $\sim\!\xi$, called the Caroli-de Gennes-Matricon (CdGM) state. The important points to note are as follows; (i) The eigenenergy of the CdGM state of the vortex core is embedded in the low energy region, $E/E_F \!=\! 2/(k_F\xi)^2$ with $k_F$ being the Fermi wave number \cite{cdgm}, and its lowest energy level is slightly shifted from the Fermi level. Also, (ii) the eigenenergy inside the core is well discretized. Hence, the spectrum of the quasiparticle states yields the particle-hole asymmetry inside the core \cite{hayashi1}. It has been found that due to the asymmetric and discretized spectrum of the CdGM state, the particle density can be suppressed inside the core \cite{hayashi2}, called the ``quantum depletion'', which makes a vortex visible via the density profile even in the Fermi system. The depletion can be gradually enhanced as the core radius $\xi$ approaches the mean inter-particle distance $k^{-1}_F$, {\it i.e.}, the quantum limit \cite{hayashi2,nygaard1,nygaard2,bulgac,mmachida1,mmachida2,chien,sensarma}. This fact is directly observed in ultracold atomic systems under Feshbach resonance \cite{zwierlein}.

The microscopic studies on multiply quantized or giant vortices have been carried out in the type-II superconductors \cite{ytanaka,volovik,virtanen,tanaka,duncan}, which indicate that the number of the branch of the CdGM state is closely associated with the topological structure of the order parameter \cite{volovik}. 
Recently, a macroscopic manifestation of giant vortices in neutral fermionic atoms was theoretically studied by Hu and Liu \cite{hu2}, who found the oscillating pattern of the particle density profiles inside the core in the population balanced system. 

The giant vortex state in ultracold Fermi atoms has not been experimentally accomplished so far. In BEC's, in contrast, the giant vortices have been experimentally created by using several techniques: The topological phase imprinting method \cite{leanhardt,shin,kumakura}, a fast rotating BEC confined in a quadratic plus quartic trap \cite{bretin}, and the transfer of the orbital angular momentum from Laguerre-Gaussian photons to a BEC \cite{phillips}. In addition, the Feshbach resonance, which controls the $s$-wave scattering length $a$, enables one to continuously connect the BEC with the Fermionic superfluid, {\it i.e.}, the BEC to Bardeen-Cooper-Schrieffer (BCS) crossover. By sweeping the magnetic field, the BEC of long-lived molecules in $a \! > \! 0$ can be transferred into the Cooper pair of fermionic atoms with the weak attractive interaction $a \!<\! 0$, across the unitary limit $a \!\rightarrow \! \pm \infty$.

In this paper, we propose a manifestation due to the topological structure of the order parameter with a multiply quantized vortex. Under population imbalance in two hyperfine spin states ($\uparrow$- and $\downarrow$-spins), the occupation difference of the CdGM state makes the core of a singly quantized vortex with the winding number $w \!=\! 1$ be magnetized \cite{takahashi,hu1}. Here, in the BCS regime $a \!<\! 0$, by solving the Bogoliubov-de Gennes (BdG) equation under population imbalance, we find that the magnetization profile inside the core of the multiply quantized vortex yields an oscillation pattern with several peaks located on a concentric circle. This oscillation is well understandable from the novel quasiparticle embedded in the vicinity of $E_F$. To help this understanding, we also derive the analytic solution for the CdGM states in multiply quantized vortices, by following the procedure which was proposed by Caroli {\it et al.} \cite{cdgm}. The analysis is extended to the vicinity of the unitary limit ($a\!\rightarrow\! \infty$) of harmonically trapped Fermi gases. Here, we discuss that in the BCS-BEC crossover, the oscillation pattern in the magnetization profile inside the core is changed from that in the BCS regime due to the strong coupling effect, which embodies the shift of the energy level of the low-lying CdGM state.

This paper is organized as follows. In Sec.~II, after introducing the theoretical framework based on the BdG equation, we present the numerical results on multiply quantized vortex states under population imbalance. This is carried out in the weak coupling regime in Sec.~II. The analysis is extended to trapped Fermi gases with population imbalance in the vicinity of Feshbach resonance in Sec.~III. Here, we discuss the crossover from the weak coupling regime to the Feshbach resonance region. Following the previous work \cite{mizushima4}, we introduce the regularization of the gap equation and the particle number equation, which allows one to describe the microscopic structure in the BCS-BEC crossover. Then, we shall show the numerical results on the macroscopic and quasiparticle structures in resonant Fermi systems with giant vortices. The final section is devoted to conclusions and discussions. In Appendix, the details on the analytical solution of the BdG equation in giant vortices shall be described. Throughout this paper, we put $\hbar\!=\!k_B\!=\!1$.

\section{Giant vortices with population imbalance: Weak coupling limit}

\subsection{Theoretical framework}

Let us consider a single vortex state with arbitrary winding number $w \!\ge\! 1$, where the pair potential can be expressed in the cylindrical coordinate ${\bm r} \!=\! (r,\theta,z)$ as
\begin{eqnarray}
\Delta ({\bm r}) = \Delta (r) e^{i w \theta}.
\label{eq:delta1}
\end{eqnarray}
Without the loss of generality, $\Delta(r)$ is the real function. We start with the BdG equation for the quasiparticle wave function $u_{\bm q}({\bm r})$ and $v_{\bm q}({\bm r})$ labeled by the quantum number ${\bm q}$ \cite{mizushima1,mizushima2,mizushima3,mizushima4,takahashi},
\begin{eqnarray}
\left[ 
\begin{array}{cc}
\mathcal{K} _{\uparrow}({\bm r}) & \Delta ({\bm r}) \\
\Delta ^{\ast} ({\bm r}) & - \mathcal{K}_{\downarrow}({\bm r})
\end{array}
\right] 
\left[ 
\begin{array}{c} u_{\bm q}({\bm r}) \\ v_{\bm q}({\bm r}) \end{array}
\right] =  E_{\bm q}
\left[ 
\begin{array}{c} u_{\bm q}({\bm r}) \\ v_{\bm q}({\bm r}) \end{array}
\right].
\label{eq:bdg}
\end{eqnarray}
The single particle Hamiltonian is given by $\mathcal{K}_{\sigma}({\bm r}) \!=\! -\frac{\nabla^2}{2M} - \mu _ \sigma + V({\bm r})$, where $M$ is the mass of fermions and $\mu _{\uparrow,\downarrow} \!=\! \mu \pm \delta \mu$ is the chemical potential of the spin $\sigma$ state. $V({\bm r})$ denotes a background potential. Here, we impose the periodic boundary condition along the $z$-axis with the period $L$. Since the resulting system yields the axisymmetry, the quasiparticle wave function may be expressed by 
\begin{eqnarray}
\left[ \begin{array}{c}
u_{\bm q} ({\bm r}) \\ v_{\bm q} ({\bm r}) 
\end{array}
\right]
= e^{iq_zz}
\left[ \begin{array}{c}
u_{\bm q}(r)e^{iq_{\theta}\theta} \\ v_{\bm q}(r)e^{i(q_{\theta} - w)\theta}
\end{array}
\right],
\label{eq:uv}
\end{eqnarray}
where the eigenstate is labeled by the quantum number ${\bm q} \!=\! (q_r,q_{\theta},q_z)$ with $q_{\theta} \!=\! 0, \pm 1, \pm 2, \cdots$ and $q_z \!=\! 0, 2\pi/L, 4\pi /L, \cdots$.

The BdG equation (\ref{eq:bdg}) is self-consistently coupled with the gap equation given by
\begin{eqnarray}
\Delta({\bm r}) = g \sum_{|E_{\bm q}| < E_c} u_{\bm q}({\bm r}) v^{\ast}_{\bm q}({\bm r}) f(E_{\bm q}),
\label{eq:gap}
\end{eqnarray} 
where $f(E _{\bm q}) \!=\! 1/(e^{E _{\bm q}/T} + 1)$ is the Fermi distribution function, $E_c$ is the energy cutoff, and $g$ is the coupling constant. Because of the breaking of the particle-hole symmetry due to the chemical potential shift $\delta \mu/2 \!\equiv\! \mu _{\uparrow} - \mu _{\downarrow}$, the sum in Eq.~(\ref{eq:gap}) is carried out for all eigenstates with both the positive and negative eigenenergies. Throughout this paper, we set $\delta\mu \!\ge\! 0$, which means the spin-up component becomes majority. 

In this section, in order to focus on the vortex core, we restrict our attention to the vortex state with the weak pair potential $\Delta _0 \!=\! 0.1E_F$ in the absence of the background potential, {\it i.e.}, $V({\bm r}) \!=\! 0$. Here $\Delta _0$ is the pair potential in the bulk at $T\!=\! 0$, and $E_F$ is the Fermi energy in an ideal Fermi gas. Then, the BdG equation (\ref{eq:bdg}) can be numerically solved by using the Bessel function expansion \cite{gygi}
\begin{eqnarray}
\left[
\begin{array}{c}
u_{\bm q}({\bm r}) \\  v_{\bm q}({\bm r})
\end{array}
\right]
= e^{iq_zz} e^{iq_{\theta}\theta} \sum^{N_B}_{i = 1} 
\left[
\begin{array}{c}
C^{(q_{\theta},q_z)}_{i} \varphi^{(q_{\theta})}_{i}(r) \\
e^{-iw\theta}D^{(q_{\theta},q_z)}_{i} \varphi^{(q_{\theta}-w)}_{i}(r)
\end{array}
\right], 
\end{eqnarray}
where the basis function,
\begin{eqnarray}
\varphi^{(\nu)}_{i}(r) \equiv \frac{\sqrt{2}}{RJ_{\nu +1}(\alpha^{(\nu)}_i)}
J_{\nu}\left(\frac{\alpha^{(\nu)}_i}{R}r \right),
\end{eqnarray} 
satisfies the orthonormal condition, $\int^{R}_0 \varphi^{(\nu)}_i(r)\varphi^{(\nu)}_{j}(r) rdr \!=\! \delta _{i,j}$. Here $J_{\nu}$ is the $\nu$-th Bessel function and $\alpha^{(\nu)}_i$ is the $i$-th zero of $J_{\nu}$. Throughout this section, all physical quantities are scaled by the length unit $k^{-1}_{F} \!\equiv\! \sqrt{2ME_F}$ and the energy unit $E_F$, and we fix the radius of the cylinder $R\!=\!200k^{-1}_{F}$, the height $L\!=\! 50k^{-1}_F$, the number of the basis function $N_B \!=\! 100$, and the chemical potential $\mu \!=\! E_F$. Note that the situation described in this section is applicable not only to ultracold Fermi atoms but also superconductors under a magnetic field, acting on electron spins.


\subsection{Quasiparticle structure and local magnetization inside giant cores}

In an isolated vortex with arbitrary winding number $w \!\ge\! 1$, the eigenenergy of the CdGM state, whose wave function is localized around the core, is analytically given from the BdG equation (\ref{eq:bdg}) as
\begin{eqnarray}
E_{\bm q} = - \left( q_{\theta}-\frac{w}{2}\right) \frac{\omega _0}{\sin(\alpha)}
+ \left( n + \frac{w-1}{2} \right) \sin(\alpha)\omega _1, 
\label{eq:cdgm}
\end{eqnarray}
with $n \!=\! 0, \pm 1, \pm 2, \cdots$ and $q_{\theta} \!=\! 0, \pm 1, \cdots$. We also introduce $\sin^2(\alpha) \!\equiv\! 1 - q^2_z/(2M\mu) $. For simplicity, from now on, we consider the eigenstates with $q_{z} \!=\! 0$, {\it i.e.}, $\alpha \!=\! \pi/2$. The two coefficients $\omega _{0,1}$ are shown in Eq.~(\ref{eq:analyticE}) and the details on the derivation of Eq.~(\ref{eq:cdgm}) are also described in Appendix. 

\begin{figure}[b!]
\includegraphics[width=0.95\linewidth]{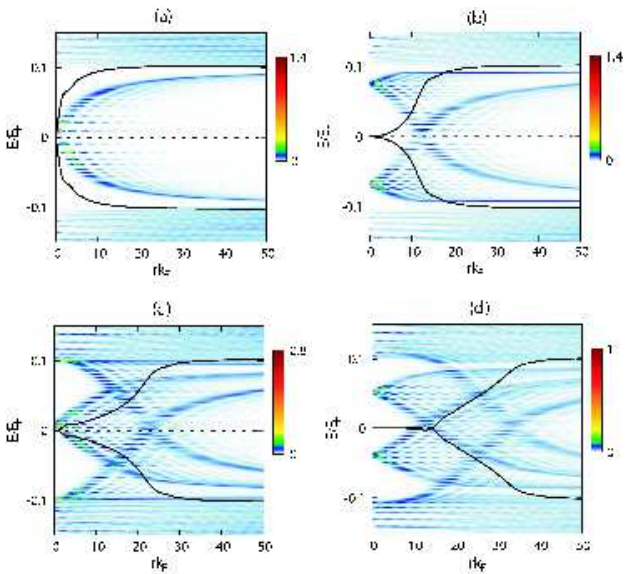} 
\caption{(Color online) Local density of states $\mathcal{N}(r,E)$ around the core in multiply quantized vortices without population imbalance in the case of $w \!=\! 1$ (a), $w\!=\!2$ (b), $w\!=\!3$ (c), and $w \!=\!4$ (d) at $T\!=\! 0$, where only the eigenstates with $q_z \!=\! 0$ are taken into account. The solid line shows the corresponding local pair potential $\pm \Delta (r)$. Throughout this paper, the Fermi surface at $\delta\mu\!=\! 0$ is put on $E\!=\! 0$.
}
\label{fig:ldos}
\end{figure}

The expression of $E_{\bm q}$ in Eq.~(\ref{eq:cdgm}) is composed by two different energy scales: (i) The energy scale comparable with the pair potential, $\omega _1 \!\sim\! \Delta _0$ and (ii) the much smaller energy scale $\omega _0 \!\sim\! \Delta^2_0/E_F \!\ll\! \omega _{1}$ in the weak coupling limit $\Delta _0 \!\ll\! E_F$. For a single vortex state with $w\!=\! 1$, the lowest branch of the dispersion relation, {\it i.e.}, $n\!=\! 0$, reproduces the well-known CdGM eigenstate with $E_{\bm q} \!\simeq\! - \frac{\Delta^2_0}{2E_F}(q_{\theta}-\frac{1}{2})$ \cite{cdgm}. The lowest CdGM state yields the ``pseudo-zero'' energy $E_{\bm q} \!\sim\! \Delta^2_0/E_F \!\ll\! \Delta _0$ in the weak coupling regime. The wave function is localized in the core region within $r\!\le\! \xi$ (see Fig.~\ref{fig:ldos}(a)), where $\xi$ is the coherence length, $\xi\!\equiv\! k_F/(M\Delta _0)$, and characterizes the core radius of the $w\!=\! 1$ vortex. In addition, Eq.~(\ref{eq:cdgm}) for arbitrary winding number $w\!\ge\! 1$ is in good agreement with the result obtained from the semiclassical approximation \cite{duncan}. 

Here, two important facts arising from Eq.~(\ref{eq:cdgm}) should be mentioned: (i) For the odd number of vorticity $w$, there always exists the core bound CdGM state with the pseudo zero energy and $q_{\theta} \!=\! 0$, because the second term in Eq.~(\ref{eq:cdgm}) can vanish. In general, the wave function is well expressed by Eq.~(\ref{eq:uv_r0}) in the region within $r\!\ll\! \xi$, {\it i.e.}, $u_{\bm q}(r) \!\sim\! J_{q_{\theta}}(k_Fr) \!\sim\! (k_Fr)^{|q_{\theta}|}$. Hence, the wave function of the lowest CdGM state with $q_{\theta} \!=\! 0$ may have a large intensity on the vortex center. In contrast, the pseudo zero ``core bound'' state with $q_{\theta} \!=\! 0$ never appears in the case of the even number $w$, because the second term in Eq.~(\ref{eq:cdgm}) remains finite. The energy of the low-lying state on the vortex center is of the order of $\Delta _0$. (ii) Apart from the core bound state with $q_{\theta} \!=\! 0$, for $w \!>\! 1$, there may exist the state with the pseudo zero energy $E_{\bm q} \!\sim\! 0$ even in the even number of $w$ when $|q_{\theta}| \!\sim\! \mathcal{O}(k_F\xi)$. In addition, the number of such state is determined by the winding number of vortices, as we will show below. Hence, it is concluded that the pseudo zero energy states can appear inside or around a giant vortex core with arbitrary winding number, while the peak position of the wave function sensitively depends on the winding number, namely, the topological structure of $\Delta({\bm r})$. 


To numerically confirm the direct relation between the winding number and the pseudo zero energy states, it is worth to visualize the quasiparticle structure of the $w$-fold quantized vortex in balanced systems. Here, we introduce the local density of states (LDOS), $\mathcal{N}_{\sigma}({\bm r},E)$, which is given by 
\begin{subequations}
\label{eq:ldos}
\begin{eqnarray}
\mathcal{N}_{\uparrow}({\bm r},E) = \sum _{\bm q} |u_{\bm q}({\bm r})|^2 \delta(E - E_{\bm q}),  
\end{eqnarray}
\begin{eqnarray}
\mathcal{N}_{\downarrow}({\bm r},E) = \sum _{\bm q} |v_{\bm q}({\bm r})|^2 \delta(E+E_{\bm q}).
\end{eqnarray}
\end{subequations} 
Figure~1 shows the LDOS for vortex states with $w\!=\! 1, 2, 3, 4$ in the case of population balance, $\delta \mu \!=\! 0$, where $\mathcal{N}({\bm r},E) \!\equiv\! (\mathcal{N}_{\uparrow}({\bm r},E) + \mathcal{N}_{\downarrow}({\bm r},E))/2$. In the core of the odd-number vortex ($w\!=\! 1, 3$), the lowest eigenstate is situated at $r \!=\! 0$ and in the vicinity of the Fermi level ($E \!\sim\! 0$). As we have mentioned above, the vortex state with the even $w$ has the distinct energy gap on the center of the vortex, where the excitation spectrum is almost symmetric with respect to the Fermi level. The CdGM branches in the giant vortex with $w\!>\! 1$ always touch the zero energy in the position far from the vortex center, {\it e.g.}, for $w\!=\! 2$, the CdGM branches in Fig.~\ref{fig:ldos}(b) become zero around $rk_F\!\simeq\! 12$, which is comparable with the coherence length $k_F\xi \!=\! 20$. In addition, the number of the pseudo zero energy state depends on the winding number of $\Delta ({\bm r})$. For instance, it is seen from Fig.~\ref{fig:ldos}(c) that the vortex state with $w\!=\! 3$ has two pseudo zero modes: The wave function of one mode has a peak at $r\!=\! 0$ and the other is at $r\!\simeq\! 25k^{-1}_F$. Hence, it is numerically confirmed that the peak position and the number of the pseudo zero energy states depends on the winding number $w$ of vortices.

\begin{figure}[t!]
\includegraphics[width=0.48\linewidth]{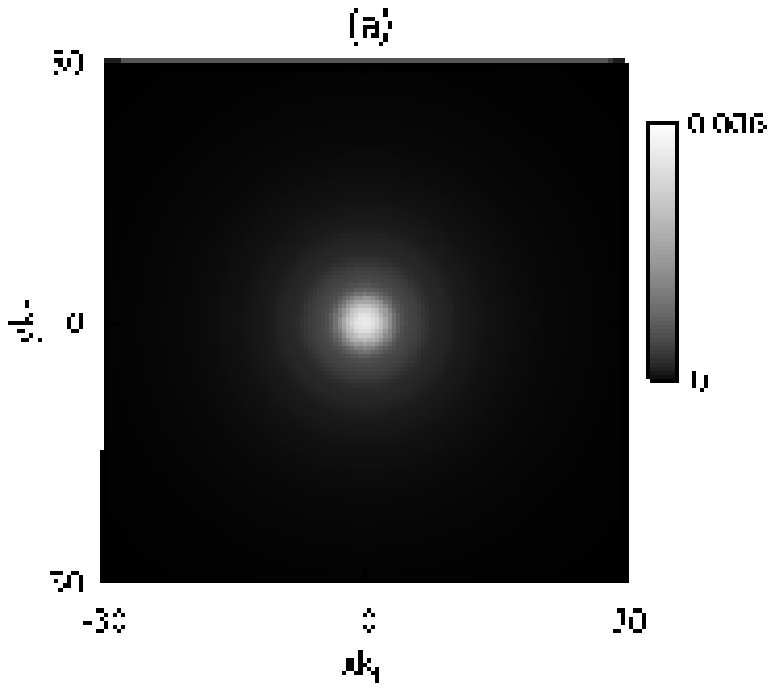} 
\includegraphics[width=0.48\linewidth]{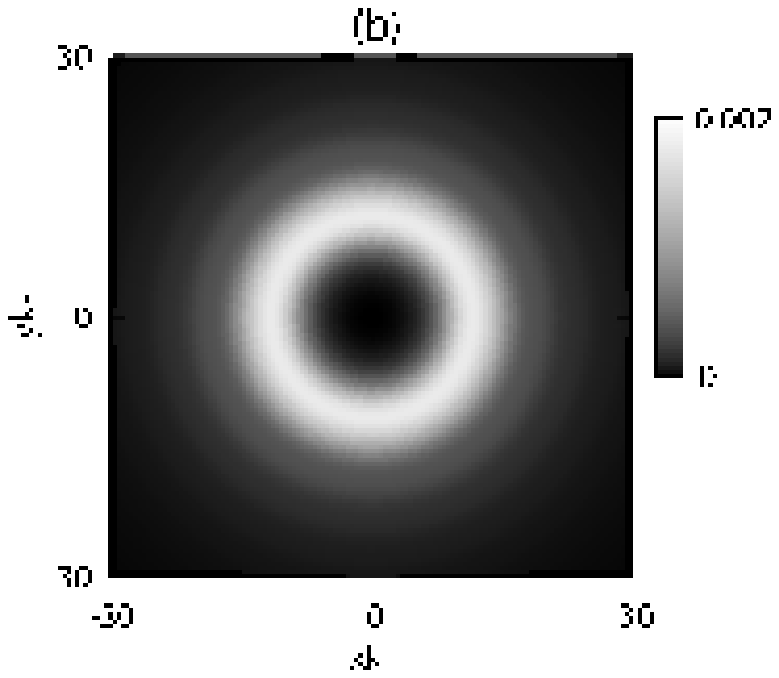} \\
\includegraphics[width=0.48\linewidth]{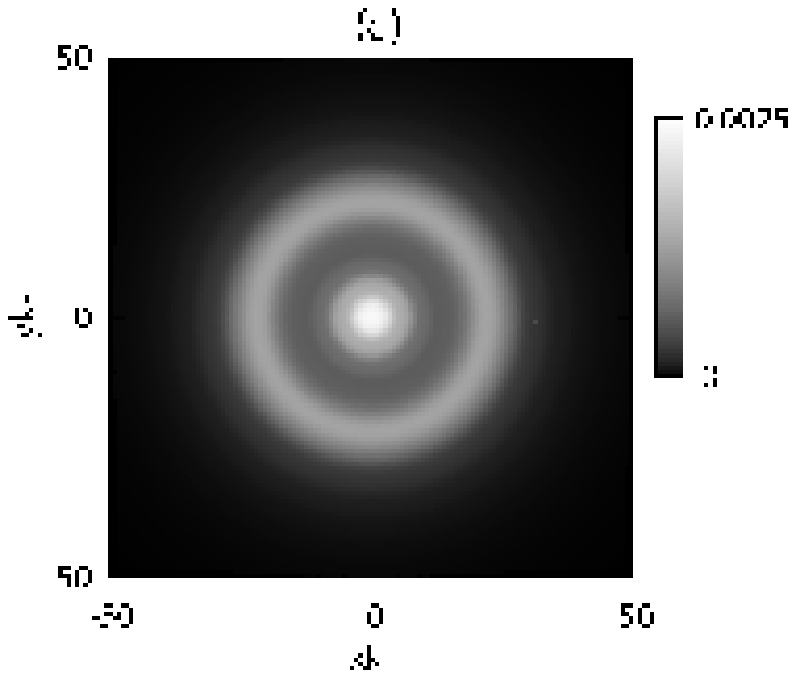} 
\includegraphics[width=0.48\linewidth]{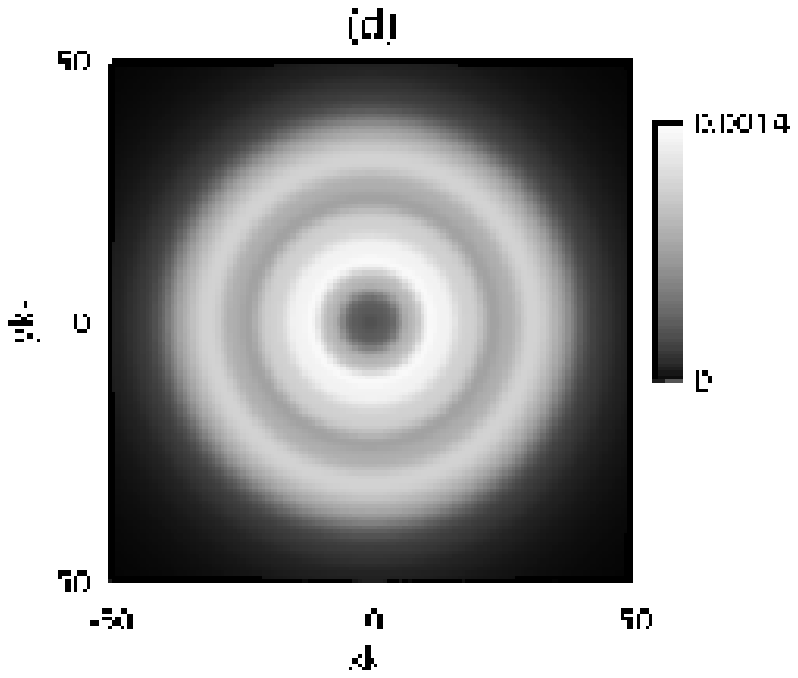} 
\caption{Local magnetization $m(x,y)$ around the vortex center in multiply quantized vortices with $w \!=\! 1$ (a), 2 (b), 3 (c), and 4 (d). All the data are at $\delta \mu \!=\! 0.02E_F$ and $T\!=\! 0$.
}
\label{fig:mag}
\end{figure}

The presence of the splitting of the Fermi level between two spin states, {\it i.e.}, $\delta \mu \!>\! 0$, enables to reveal the two characteristic features of the CdGM states inside the giant vortex core, namely, (i) the pseudo-zero or gapful excitation at the core $r\!=\! 0$ and (ii) the ``pseudo'' gapless excitation at $r \!\sim\! \xi$. Figure~\ref{fig:mag} shows the local ``magnetization'' profile, 
\begin{eqnarray}
m({\bm r}) \equiv \rho _{\uparrow}({\bm r}) - \rho _{\downarrow}({\bm r}),
\end{eqnarray} 
around the core region of giant vortices with $w\!=\! 1, 2, 3, 4$. Here, $\rho _{\uparrow,\downarrow}({\bm r})$ are the local particle densities in hyperfine spin states, which are given by
\begin{subequations}
\label{eq:defrho}
\begin{eqnarray}
\rho _{\uparrow}({\bm r}) = \sum _{\bm q} |u_{\bm q}({\bm r})|^2 f(E_{\bm q}), 
\label{eq:rhou} 
\end{eqnarray}
\begin{eqnarray}
\rho _{\downarrow}({\bm r}) = \sum _{\bm q} |v_{\bm q}({\bm r})|^2 \left[1- f(E_{\bm q})\right]. 
\label{eq:rhod} 
\end{eqnarray}
\end{subequations}

The magnetization of the vortex core results from the fact that the excess atoms of the majority component can be accommodated by the eigenstates with the pseudo-zero energy. For instance, the core of the single quantized ($w\!=\! 1$) vortex is filled in by the paramagnetic moment, as seen in Fig.~\ref{fig:mag}(a). The accommodation of the magnetic moment inside the core is associated with the low-lying quasiparticle state, {\it i.e.}, the CdGM state. Since $\rho _{\uparrow,\downarrow}({\bm r}) \!=\! \int^{\infty}_{-\infty} \mathcal{N}_{\uparrow,\downarrow}({\bm r},E)f(E)dE \!\simeq\! \int^{E^{(\uparrow,\downarrow)}_F}_{-\infty} \mathcal{N}({\bm r},E)dE$ at $T\!=\! 0$, the local magnetization $m({\bm r})$ is determined by the eigenstates embedded in the spacing between the Fermi surfaces of two spin species. Here, $\mathcal{N}({\bm r},E)$ is the LDOS at $\delta\mu \!=\! 0$, shown in Fig.~\ref{fig:ldos}. In the case of imbalanced population, the Fermi energy of the majority (minority) spin component is shifted from that of the balanced case upward (downward), $E^{(\uparrow)}_F \!=\! E_F + \delta\mu$ ($E^{(\downarrow)}_F \!=\! E_F -\delta\mu$). Hence, the low-lying CdGM states are embodied by the local paramagnetic moment inside the core, when $\delta\mu \!=\! 0$.

For the odd-number vortex, since the lowest CdGM state at the vortex center has the much small energy $\sim\! \Delta^2_0/E_F$, the vortex center is always magnetized in the presence of the small splitting of the Fermi level, $\delta\mu \!\ll\!\Delta _0$, as seen in Figs.~\ref{fig:mag}(a) and (c). This is contrast to that in the even number vortex, where the vortex center can not be magnetized in the situation of $\delta\mu \!\ll\!\Delta _0$ because of the large energy gap of the CdGM state. Furthermore, it is seen from Figs.~\ref{fig:mag}(b)-(d) that the pseudo gapless excitation at $r \!\neq\! 0$ leads to the concentric oscillation pattern of the local magnetization, depending on the winding number $w$. For instance, in the case of the doubly quantized vortex with $w\!=\! 2$ in Fig.~\ref{fig:ldos}(b), since the branches of the CdGM state touch the Fermi level at $r\!=\! 12k^{-1}_F$ and their excitations have almost zero energy, the excess atoms can be accommodated there under small splitting of the Fermi level $\delta\mu \!\ll\!\Delta _0$. The peak position of $m({\bm r})$ at $r \!\sim\! 12 k^{-1}_F$ in Fig.~\ref{fig:mag}(b) coincides to the spatial profile of the pseudo zero states in Fig.~\ref{fig:ldos}(b). The $w \!=\! 4$ vortex has two crossing points of the CdGM branches on the Fermi level as seen in Fig.~\ref{fig:ldos}(d), leading to the double peak structure on the concentric circles with radius $rk_F \!\sim\! 15$ and 30 in Fig.~\ref{fig:mag}(d).

\begin{figure}[b!]
\includegraphics[width=0.95\linewidth]{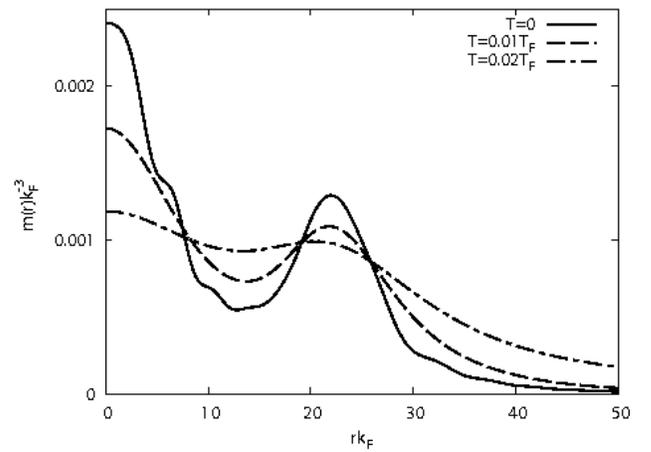} 
\caption{Local magnetization around the core region of the $w\!=\! 3$ vortex within $r \!<\! 50k^{-1}_F$ at $T/T_F \!=\! 0$ (solid), $0.01$ (dashed), $0.02$ (dashed-dotted line). The chemical potential shift is fixed as $\delta\mu \!=\! 0.02E_F$ and $\max|\Delta(r)| \!=\! 0.1E_F$ at $T\!=\! 0$. }
\label{fig:tmp}
\end{figure}

Before turing to the unitary limit, we mention to the temperature dependence of the local magnetization inside the giant vortex core. Figure~\ref{fig:tmp} shows the local magnetization around the core region $r\!\le\! 50 k^{-1}_F$ of the $w\!=\!3$ vortex at several temperatures, $T/T_F\!=\! 0, 0.01, 0.02$, where $\delta\mu \!=\! 0.02E_F$ is fixed. It is found that the temperature comparable with $\delta\mu$ smoothes the concentric peak structure in $m(r)$. In particular, we should emphasize that the paramagnetic moment in the outside of the region $r\!>\! 30k^{-1}_F$ increases as $T$ increases, which is characterized by the Yosida function \cite{yosida}, while that at the vortex center ($r\!=\! 0$) and at the concentric peak ($r\!\simeq\! 25k^{-1}_F$) decreases.

\section{Giant vortices with population imbalance: trapped Fermi gases near Feshbach resonance}

\subsection{BCS-BEC crossover theory}

Having obtained the direct relation between the quasiparticle structures and magnetization inside giant vortex cores in the weak coupling limit, let us now proceed to extend the analysis into the more realistic situation, such as a trapped Fermi gas under an $s$-wave Feshbach resonance. Here, we consider the three dimensional cylindrical system that the fermions are confined by the two dimensional trap potential $V({\bm r}) \!=\! \frac{1}{2}M\omega^2r^2$, where $r^2 \!\equiv\! x^2 + y^2$. To do with the vicinity of $s$-wave Feshbach resonance, we have to modify the theoretical framework based on the BdG equation in Sec.~II A. First, the ultra-violet divergence in Eq.~(\ref{eq:gap}) is removed by replacing $g$ to the effective coupling constant $\tilde{g}$, which is associated with the energy cutoff $E_c$ and the dimensionless coupling constant $1/(k_Fa)$ \cite{sademelo,sensarma},
\begin{eqnarray}
\frac{E_F}{\tilde{g}k^3_F} = \frac{1}{8\pi k_Fa} +\frac{1}{4\pi^2}\sqrt{\frac{E_c}{E_F}},
\label{eq:geff}
\end{eqnarray}
where $a$ is an $s$-wave scattering length. Also, the chemical potential $\mu$ is adjusted during the iteration to conserve the total particle number, 
\begin{eqnarray}
N = N_{\uparrow} + N_{\downarrow}, \hspace{3mm} 
N_{\uparrow,\downarrow}=  \int \rho _{\uparrow,\downarrow}({\bm r}) d{\bm r}, 
\label{eq:number}
\end{eqnarray}
where the particle density of each spin state is given by Eq.~(\ref{eq:defrho}). 

The BdG equation (\ref{eq:bdg}) is now self-consistently coupled with the particle number equation (\ref{eq:number}) and the regularized gap equation (\ref{eq:gap}), where the bare coupling constant $g$ in Eq.~(\ref{eq:gap}) is replaced to the effective one $\tilde{g}$ in Eq.~(\ref{eq:geff}). The set of equations is free from the energy cutoff $E_c$ and allows one to describe the qualitative feature of $T\!=\! 0$ superfluid phases in the entire range of $(k_Fa)^{-1}$ from the BCS ($(k_Fa)^{-1} \!\rightarrow\! -\infty$) to the BEC limit ($(k_Fa)^{-1} \!\rightarrow\! +\infty$). 

We numerically solve the gap equation (\ref{eq:gap}) up to the energy $E^{({\rm BdG})}\!=\! 100\omega _r$, by using the quasiparticle wave function obtained from Eq.~(\ref{eq:bdg}). In addition, the higher energy contribution above $E^{({\rm BdG})}$ is supplemented by employing the local density approximation (LDA). The details of the BdG-LDA hybrid method are described in Ref.~\cite{mizushima4}. For all data, the total particle number is conserved as $N\!=\! 9,000$, and we also set $L \!=\! 3 d$, corresponding to the Fermi energy $E_F \!=\! \omega(30\pi Nd/(16L))^{2/5}\!=\! 50 \omega$. Here, the trap length and energy scales are introduced, {\it i.e.}, $d \!\equiv\! \sqrt{1/M\omega}$ and $\omega$, respectively. The calculation is carried out in the range from the weak coupling regime $(k_Fa)^{-1} \!=\! - 1.2$ to the unitary limit $(k_Fa)^{-1} \!=\! 0$.

\subsection{Quasiparticle structure and local magnetization inside giant cores}

\begin{figure}[b!]
\includegraphics[width=0.96\linewidth]{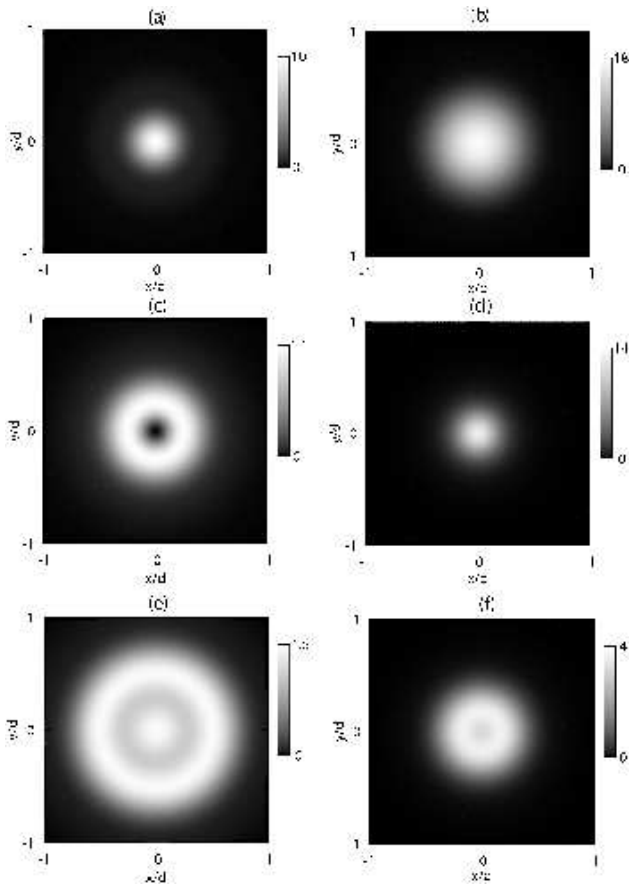} 
\caption{Local magnetization $m(x,y)$ around the vortex core with the winding number $w \!=\! 1$ (a, b), 2 (c, d), 3 (e, f) at $T\!=\! 0$. Figures (a), (c), and (e) are in the BCS side of the resonance $1/(k_Fa) \!=\! -0.4$, and the others are at the resonance $1/(k_Fa) \!=\! 0$. Population imbalance is estimated as (a) $P \!=\! 0.12$ ($w\!=\! 1$), (b) 0.82 ($w\!=\! 1$), (c) 0.014 ($w\!=\! 2$), (d) 0.03 ($w \!=\! 2$), (e) 0.014 ($w \!=\! 3$), (f) 0.03 ($w\!=\! 3$).
}
\label{fig:unitary}
\end{figure}

Here, since in the actual experiment the net magnetization is conserved due to the absence of the spin relaxation process, it is convenient to introduce the population imbalance between two hyperfine spin states,
\begin{eqnarray}
P \equiv \frac{N_{\uparrow} - N_{\downarrow}}{N_{\uparrow} + N_{\downarrow}}.
\end{eqnarray}
Similarly to the non-trapped system in Fig.~\ref{fig:mag}, it is found that the distinct pattern of the local ``magnetization'' $m({\bm r})$ for vortices with arbitrary winding number appears inside the core in the weak coupling regime $1/(k_Fa) \!=\! - 0.4, -1.2$, even in the presence of the trap potential. Note that the coherence length is estimated as $\xi\!\equiv\! k_F/(M\Delta _0) \!=\! 5.2 k^{-1}_F \!=\! 0.52d$ at $1/(k_Fa) \!=\! -0.4$ and $\xi \!=\! 16.7 k^{-1}_F \!=\! 0.167 d$ at $1/(k_Fa) \!=\! -1.2$, where $\Delta _0 \!\equiv\! \max|\Delta({\bm r})|$ at $T\!=\! 0$. The magnetization profiles inside the $w\!=\! 1$, 2, 3 vortex cores at $1/(k_Fa) \!=\! -0.4$ are displayed in Figs.~\ref{fig:unitary}(a), \ref{fig:unitary}(c), and \ref{fig:unitary}(e), respectively, where the fermions are confined by the harmonic trap. 

As the interaction approaches the unitary limit, however, the profile of $m({\bm r})$ is drastically changed. Figures~\ref{fig:unitary}(b), \ref{fig:unitary}(d), and \ref{fig:unitary}(f) show the profile of the local magnetization around the giant vortex core at the unitary limit, $1/k_Fa \!=\! 0$, where the coherence length $\xi \!=\! 3.0 k^{-1}_{F} \!=\! 0.3 d$. In the vortex state with $w\!=\!1$ at $1/(k_Fa) \!=\! 0$, the lowest CdGM state has a large energy gap comparable with $\Delta _0$, because of $\Delta _0/E_F \!=\! \mathcal{O}(1)$. Hence, the energy gap of the lowest CdGM state bounded on the vortex center leads to the absence of the magnetization inside the core even in the high value of $P$. Note that the large population imbalance in the $w\!=\! 1$ vortex in Fig.~\ref{fig:unitary}(b) results from the magnetization accommodated around the edge region of the cloud, $r \! \gg\! \xi$. It means that the volume of the magnetization $m(x,y)$ inside the core within $r/d \!<\! 1$ is almost same in all of Figs.~\ref{fig:unitary}(a)-(f).

At $1/(k_Fa) \!=\! 0$, in the vortex state with $w \!>\! 1$, we also find the completely different behavior of $m({\bm r})$ from that at $1/(k_Fa) \!=\! -0.4$. For instance, in contrast to the pattern shown in Figs.~\ref{fig:mag}(b) and \ref{fig:unitary}(c), the core of the doubly quantized vortex in Fig.~\ref{fig:unitary}(d) is filled in by a large amount of the excess fermions, which yields the single peak structure around the vortex center. Similarly, the small depletion of $m({\bm r})$ appears at $r\!=\! 0$ in the $w\!=\! 3$ vortex on a resonance, which is contrast to that in Figs.~\ref{fig:mag}(b) and \ref{fig:unitary}(c).

\begin{figure}[t!]
\includegraphics[width=0.9\linewidth]{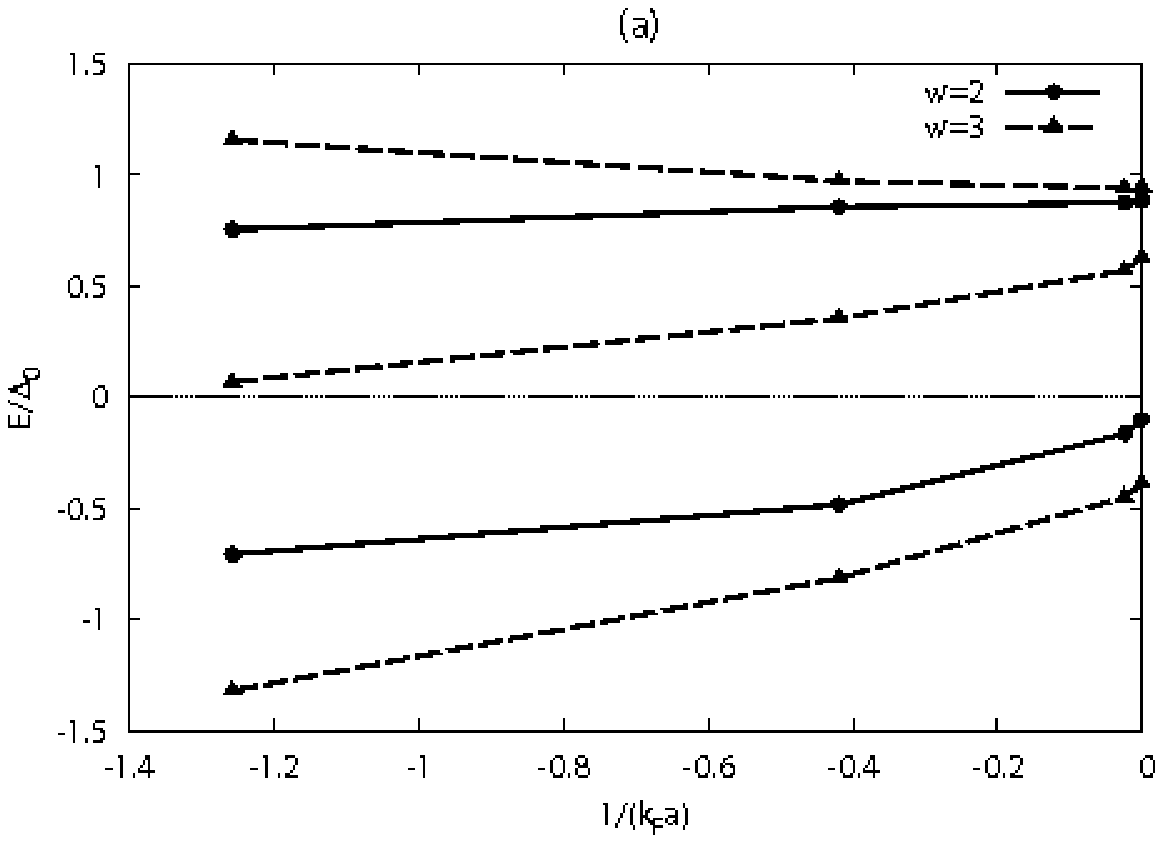} 
\includegraphics[width=0.9\linewidth]{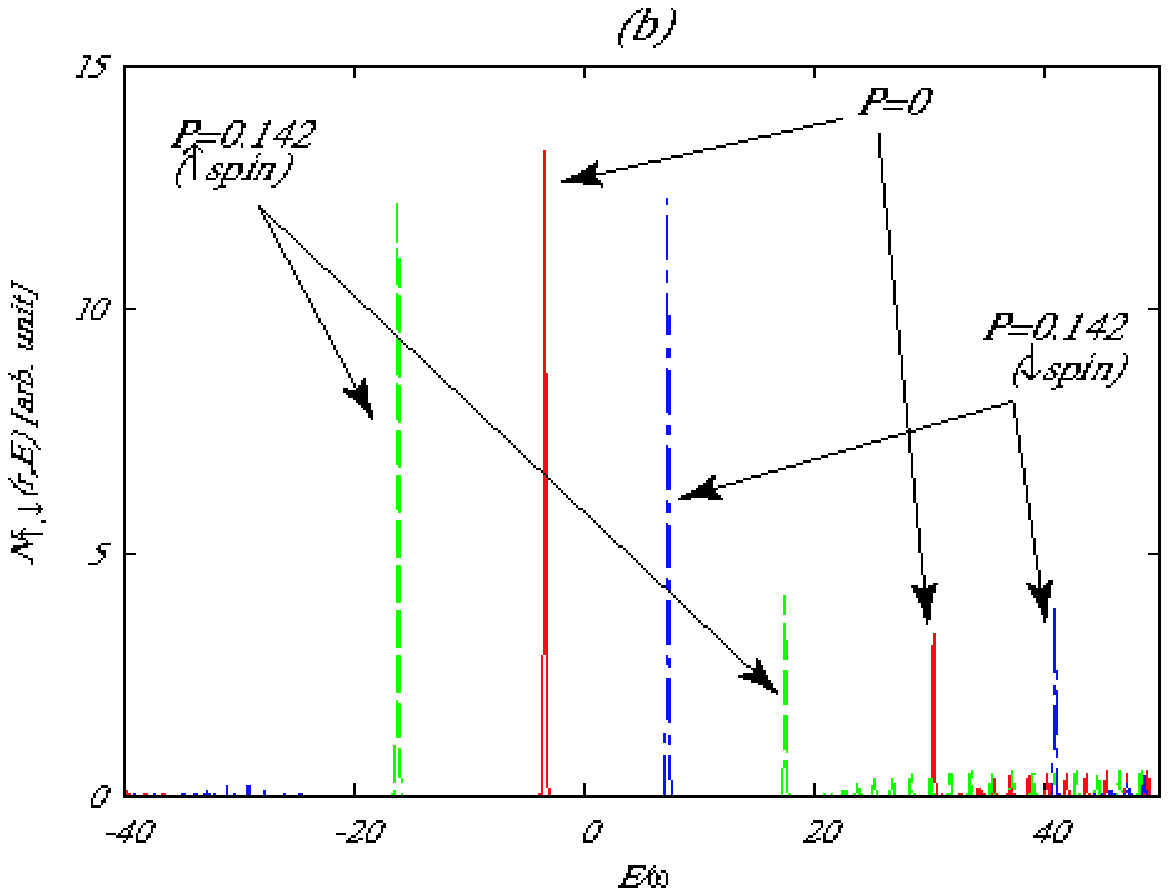} 
\includegraphics[width=0.9\linewidth]{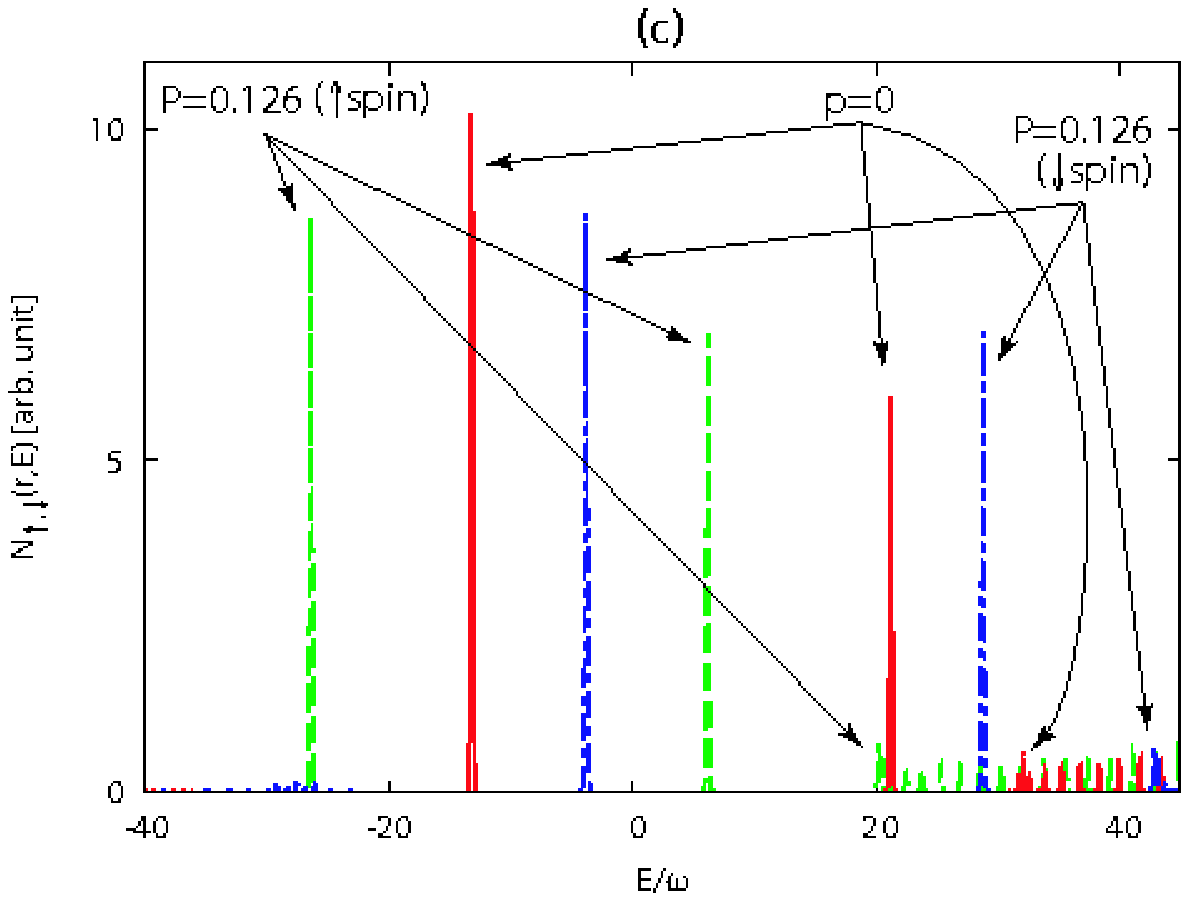} 
\caption{(a) Eigenenergies of the CdGM states having the intensity at $r \!=\! 0$ as functions of $1/(k_Fa)$. The solid and dashed lines denote the CdGM branch in the $w\!=\! 2$ and $3$ vortex states, respectively. The eigenenergies in (a) are scaled by $\Delta _0$, which is given by $\Delta _0 \!\equiv\! \max{|\Delta({\bm r})|}$. LDOS at $r\!=\! 0$, $N_{\uparrow,\downarrow}(r,E)$, of $w\!=\! 2$ (b) and $3$ (c) vortex state with and without population imbalance. The all dates in (b) and (c) are at $1/k_Fa \!=\! 0$.
}
\label{fig:ldos_uni}
\end{figure}

The behavior of $m({\bm r})$ in the core region of various giant vortices results from the strong coupling effect with $\Delta _0 /E_F\!\sim\!\mathcal{O}(1)$ at $1/k_Fa \!=\! 0$. To understand this, in Fig.~\ref{fig:ldos_uni}(a), we first plot the energy shift of the peak of $\mathcal{N}(r \!=\! 0,E)$ at $\delta \mu \!=\! 0$, as functions of $1/(k_Fa)$. As seen in Fig.~\ref{fig:ldos}(b), in the weak coupling regime with $\Delta _0/E_F \!=\! 0.1$ and $\mu \!=\! E_F$, the LDOS's $\mathcal{N}_{\uparrow,\downarrow}(r\!=\! 0, E)$ of the $w \!=\! 2$ vortex yield the double peak structures at $E \!\simeq\! \pm 0.7\Delta _0$ on the vortex center, whose spectral evolution is almost symmetric with respect to the Fermi level $E\!=\! 0$. At $1/k_Fa \!=\! 0$, however, the double peaks of $\mathcal{N}_{\uparrow,\downarrow}(r\!=\! 0, E)$ in the $w\!=\! 2$ vortex are shifted upward and their positions become asymmetric with respect to the Fermi energy, as seen in Fig.~\ref{fig:ldos_uni}(a) with the solid line. Here the lower branch stays nearby the $E\!\sim\! 0$ region in the $1/(k_Fa) \!\sim\! 0$ region. This shift is understandable from the analytic expression of the CdGM state shown in Eq.~(\ref{eq:cdgm}),
\begin{eqnarray}
E^{(w=2)}_{\bm q} = \omega _0 \pm \frac{1}{2} \omega _1,
\label{eq:cdgm2}
\end{eqnarray}
with $q_{\theta}\!=\! 0$, $q_{z} \!=\! 0$, and $n\!=\! 0, -1$. Here, let us recall that $\omega _0$ and $\omega _1$ are of the order of $\frac{\Delta^2_0}{E_F}$ and $\Delta _0$, respectively, {\it i.e.}, $\omega _0 \!\ll\! \omega _1$ in the weak coupling regime. Then, the eigenenergies of two lowest CdGM states in the $w\!=\! 2$ vortex are symmetric with respect to the Fermi level $E\!=\! 0$. Indeed, as plotted at $1/(k_Fa)\!=\! -1.2$ in Fig.~\ref{fig:ldos_uni}(a), the point below (above) $E\!=\! 0$ corresponds to the plus (minus) sign of Eq.~(\ref{eq:cdgm2}). As approaching the unitary limit $1/k_Fa \!=\! 0$, however, $\omega _0$ becomes comparable with $\omega _1$, which causes the energy shift of the CdGM state with $\omega _0 -\frac{1}{2}\omega _1$ located inside the Fermi level toward $E \!=\! 0$, {\it e.g.}, $E\!=\! -0.05 \Delta _0$ at $1/(k_Fa) \!=\! 0$ in Fig.~\ref{fig:ldos_uni}(a). This energy shifts of the CdGM states localized on the center of giant vortices are also confirmed in Ref.~\cite{hu2}, where the CdGM states shift up across $E\!=\!0$ as further approaching the BEC limit ($1/(k_Fa) \!\rightarrow\! +\infty$). 

Figure~\ref{fig:ldos_uni}(b) shows the LDOS at $r\!=\! 0$ in $w\!=\! 2$ vortex state with and without population imbalance at $1/k_Fa \!=\! 0$. In a finite population imbalance, since the energy of two spin states is shifted downward ($\uparrow$-spins) or upward ($\downarrow$-spins), the lower branch of the two CdGM states is occupied (unoccupied) by the majority (minority) spin component, which gives rise to the magnetization of the vortex center even in the even number winding vortex. 

Figure~\ref{fig:ldos_uni}(a) also tells us that as $1/(k_Fa)$ approaches the unitarity, the pseudo zero eigenstate, which appears in the $w\!=\! 3$ vortex at $1/(k_Fa) \!=\! -1.2$, is quickly shifted upward, {\it e.g.}, $E \!=\! 20 \omega \!=\! 0.57 \Delta _0$ at the unitarity, as seen in Fig.~\ref{fig:ldos_uni}(a) and \ref{fig:ldos_uni}(c). It indicates that at $1/k_Fa\!=\! 0$, the distinct energy gap may emerge in the vicinity of $E\!=\! 0$ in the case of the odd number $w$. This energy gap leads to the suppression of $m({\bm r})$ at $r\!=\! 0$ as seen in Fig.~\ref{fig:unitary}(f). This is contrast to the case of the weak coupling regime. For instance, as seen in Fig.~\ref{fig:unitary}(e), the magnetization in the $w\!=\! 3$ vortex at $1/(k_Fa)\!=\! -0.4$ has a distinct peak at $r\!=\! 0$ and also the other peak on the concentric circles. This oscillation pattern of $m({\bm r})$ becomes clear as further approaching the weak coupling limit ($1/(k_Fa) \!\rightarrow \! - \infty$).

\begin{figure}[t!]
\includegraphics[width=0.9\linewidth]{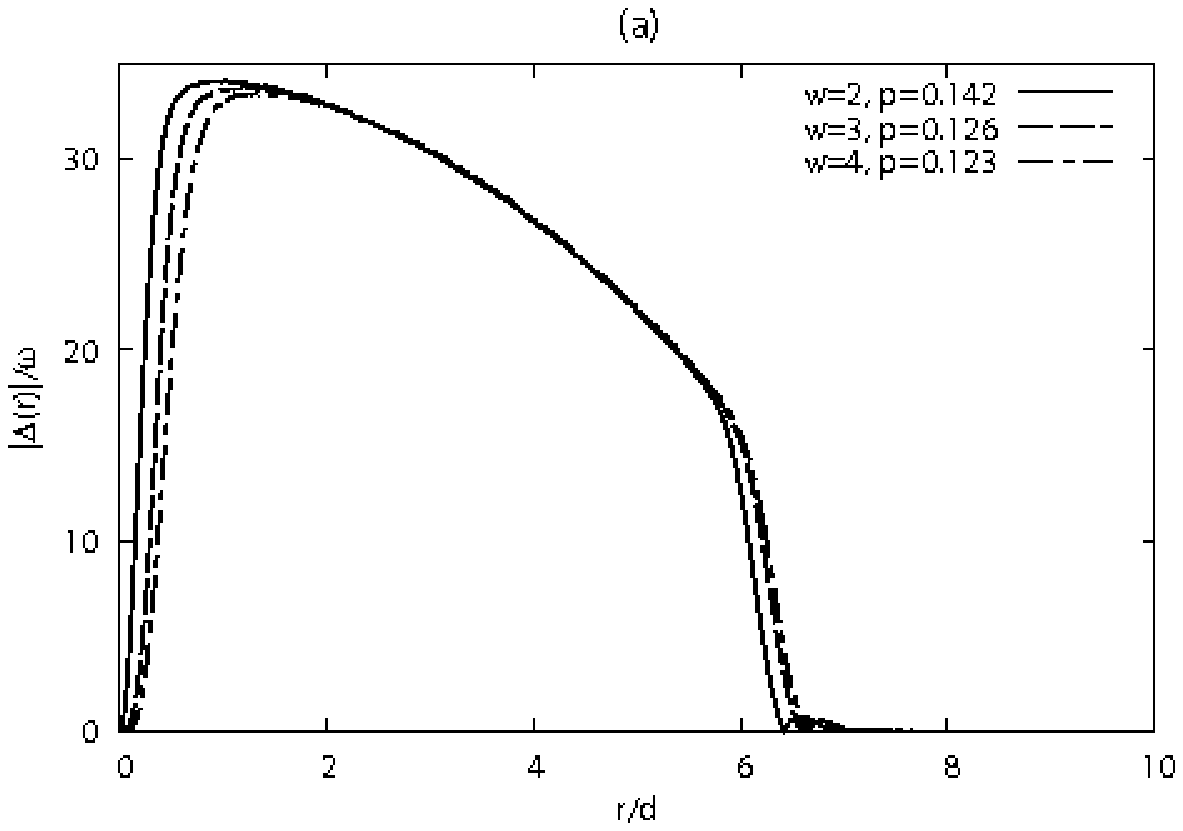} 
\includegraphics[width=0.9\linewidth]{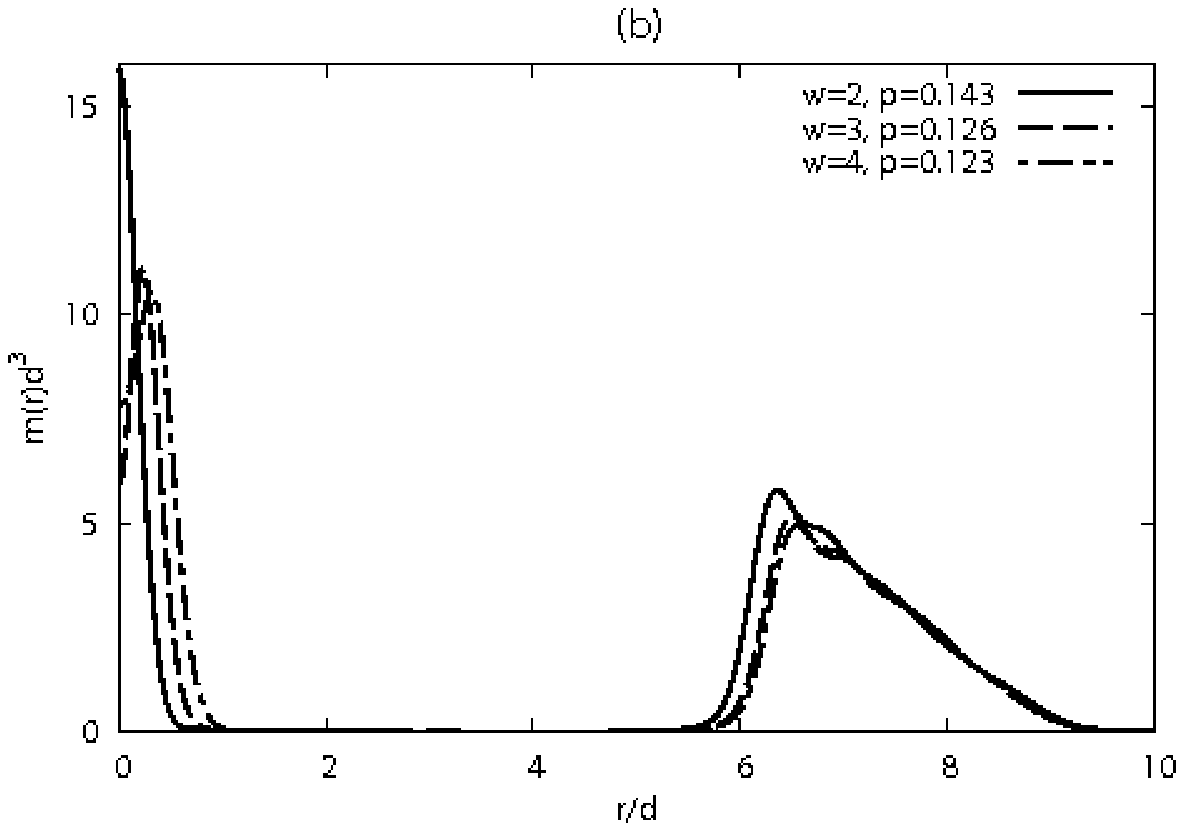} 
\caption{Profile of the local pair potential $\Delta(r)$ (a) and magnetization $m(r)$ (b) at the unitary limit, $1/k_Fa \!=\! 0$. Population imbalance is estimated in vortex states with various winding number as $P \!=\! 0.143$ ($w\!=\! 2$, solid line), 0.126 ($w\!=\! 3$, dashed line), 0.123 ($w\!=\! 4$, dashed-dotted line). 
}
\label{fig:unitary2}
\end{figure}

In addition, we mention that due to the presence of the trap potential, finite population imbalance occurs the depairing in the vicinity of the edge, in addition to the inner region of the core. The numerical results for the pairing field and local magnetization shown in Fig.~\ref{fig:unitary2} reveals the phase separated state between the superfluid and spin polarized normal domains at $1/k_Fa \!=\! 0$. Similarly to the non-vortex \cite{mizushima4} and the case of singly quantized vortex \cite{takahashi}, the pairing field in $1/k_Fa \!<\! 0$ yields the oscillation in the surface region, that is, the Fulde-Ferrell-Larkin-Ovchinnikov-like oscillation. It is found that the oscillation of $\Delta(r)$ is proper to imbalanced systems both with and without a vortex line in the extensive range $1/k_Fa \!<\! 0$. As the interaction touches the unitary limit, however, the coexistence area becomes narrow, since the oscillation period is reduced to the interparticle distance $\sim\! k^{-1}_F$ \cite{mizushima4}.

\section{Concluding remarks}

In this paper, we have investigated the core structure of multiply quantized vortex in imbalanced Fermi systems with and without a trap potential. In conclusion, it is found that in the weak coupling regime, as shown in Fig.~\ref{fig:mag}, the local magnetization inside the core of multiply quantized vortices yields the concentric oscillation pattern, which reveals the quasiparticle structure bound inside the core, called the CdGM states. The ``pseudo-zero'' energy of the CdGM states in vortices with an odd-winding number enables the accommodation of the magnetization on the vortex center. In contrast, the magnetic moment is excluded from the vortex center in the case of an even winding number. This is because the CdGM state has a distinct energy gap of the order of $\Delta _0$ as shown analytically in Eq.~(\ref{eq:cdgm}) and numerically in Fig.~\ref{fig:ldos}. 

This difference is alternatively understandable from the $\pi$-phase shift of the pairing field \cite{machida,mizushima2,ichioka,machida2,machida3}. In general, in the presence of the kink structure of the pair potential, quasiparticles across the domain wall feel the sign change of the pair potential, giving rise to the eigen modes with the zero energy, called the Andreev bound state \cite{kashiwaya} or the mid-gap state \cite{machida,machida2,machida3}. The modes are strongly bounded around the domain wall. In the case of giant vortices with an odd winding number, since quasiparticles tracing the path across the vortex center $r\!=\! 0$ always experience the $\pi$-phase shift of $\Delta ({\bm r})$, the bound state, called the CdGM state with the pseudo-zero energy in the text, appears in the vicinity of the Fermi level, whose energy scale is much smaller than the energy gap in the bulk region. In contrast, the pseudo-zero modes never appear in case of an even winding number, where the pair potential dose not change its sign along the path across the vortex center. The relation between the topological structure of $\Delta({\bm r})$ and the quasiparticle state has been shown analytically in Eq.~(\ref{eq:cdgm}) and numerically in Fig.~\ref{fig:ldos}. The pseudo-zero CdGM states can easily participate the magnetization, when the Fermi level of two spin states is mismatched. The peak structure of the local magnetization appears at $r\!=\!0$ in Fig.~\ref{fig:mag}(a) and (c). In the even number of winding, however, the distinct energy gap arising from the topological reason prevents the accommodation of the excess spins at $r\!=\!0$, as seen in Fig.~\ref{fig:mag}(b) and (d). 

In contrast to the weak coupling regime, where the magnetization inside vortex cores can be closely associated with the topological structure of the pairing field, it has been found that the magnetization profile inside the core of various giant vortices is drastically changed in the vicinity of Feshbach resonance. The key factor is that two energy scales in Eq.~(\ref{eq:cdgm}), $\omega _0$ and $\omega _1$, becomes comparable with each other as $1/k_Fa$ approaches the unitary limit. The upward shit of all core bound states gives rise to the drastic change of the magnetization profile inside cores from that in the BCS limit. 

Finally, we should emphasize that the oscillating pattern of the magnetization inside the giant core provides the spectroscopy of the discretized core structures. As seen in Fig.~\ref{fig:unitary}, the magnetization profile inside the core is drastically changed as varying $1/(k_Fa)$. Namely, it is proposed that the density experiments using phase contrast imaging reveal the spectrum of the core localized CdGM states via the magnetization.

\section*{ACKNOWLEDGMENTS}

This research is supported by the Grant-in-Aid for Scientific Research, Japan Society for the Promotion of Science.

\appendix

\section{Analytic solution of CdGM states in giant vortices}

In this Appendix, we describe the details on the derivation of the analytic expression of the CdGM states in Eq.~(\ref{eq:cdgm}), from the BdG equation (\ref{eq:bdg}) without the chemical potential shift, $\delta \mu \!=\! 0$, and trap potential $V({\bm r}) \!=\! 0$. Using Eqs.~(\ref{eq:delta1}) and (\ref{eq:uv}) and assuming $q_{z}\!\equiv\! k_{\mu}\sqrt{1-\sin^2(\alpha)} \!\ll\! k_{\mu} \!\equiv\! \sqrt{2M|\mu|^2}$, the BdG equation (\ref{eq:bdg}) can be rewritten in the cylindrical coordinate to
\begin{eqnarray}
\hspace{-3mm}
\left[ \mathcal{L}_m \hat{\tau}_0 - 2iM\Delta\hat{\tau}_2 \right] {\bm u}_{\bm q} 
= \hat{\tau}_3\left[ \frac{w(q_{\theta}-\frac{w}{2})}{r^2} - 2ME_{\bm q}\right]{\bm u}_{\bm q}.
\label{eq:bdg2}
\end{eqnarray}
Here we set $\Delta \!\equiv\! \Delta(r)$, ${\bm u}_{\bm q} \!\equiv\! {\bm u}_{\bm q}(r) \!=\! [u_{\bm q}(r),v_{\bm q}(r)]^T$, and 
\begin{eqnarray}
\mathcal{L}_m \!\equiv\! \frac{d^2}{dr^2} + \frac{1}{r}\frac{d}{dr} - \frac{m^2}{r^2} + k^2_{\mu}\sin^2(\alpha)
\end{eqnarray} 
with $m \!=\! \sqrt{q^2_{\theta}-wq_{\theta}+\frac{w^2}{2}}$. Also, we use the Pauli matrices $\hat{\tau}_{1,2,3}$ and $2\!\times\!2$ unit matrix $\hat{\tau}_0\!=\! {\rm diag}(1,1)$. Throughout this Appendix, we consider the BCS regime with $\mu \!>\! 0$. 

Following the procedure proposed by Caroli {\it et al.} \cite{cdgm}, we introduce a radius $r_c$ that $\Delta(r) \!=\! 0$ for $r\!<\!r_c$. Then, the BdG equation (\ref{eq:bdg2}) can be analytically solved if one assume the following conditions: (i) $|q_{\theta}| \!\ll\! r_ck_F \!\ll\! k_F\xi$, (ii) $E_{\bm q} \!\ll\! \Delta _0$, and (iii) $E_{\bm q} \!\ll\! |\mu|^2\sin^2(\alpha)$. 

The wavefunction in Eq.~(\ref{eq:bdg2}) is obtained in the range $r \!<\! r_c$ as 
\begin{eqnarray}
{\bm u}_{\bm q}(r) = 
\left[
\begin{array}{c}
A_u J_{q_{\theta}}(k_+(\alpha) r) \\ A_v J_{q_{\theta}-w} (k_-(\alpha)r)
\end{array}
\right],
\label{eq:uv_r0} 
\end{eqnarray} 
where $A_u$ and $A_v$ are the arbitrary constant and we set
\begin{eqnarray}
k_{\pm}(\alpha) \!\equiv\! k_{\mu}\sin(\alpha)\pm \frac{E_{\bm q}}{v_{\mu}(\alpha)},
\end{eqnarray}
with $v_{\mu}(\alpha) \!=\! k_{\mu}\sin(\alpha)/M$. 

For $r\!>\! r_c$, the wave functions are composed of the Hankel function $H^{(i)}_m$ and the slow functions ${\bm \varphi}_i(r)$ varying over the order of $\xi$ \cite{cdgm},
\begin{eqnarray}
{\bm u}_{\bm q}(r) = \sum _{i=1,2} H^{(i)}_m (k_{\mu}\sin(\alpha)r){\bm \varphi}_i(r).
\label{eq:hankel}
\end{eqnarray}
Then, Eq.~(\ref{eq:bdg2}) can be reduced to 
\begin{eqnarray}
\left[ \hat{\tau}_0 \frac{d}{dr} - \hat{\tau}_2\frac{\Delta}{v_{\mu}} \right]
{\bm \varphi}_1 = i\hat{\tau}_3 
\left[ \frac{E_{\bm q}}{v_{\mu}} - \frac{w(q_{\theta}-\frac{w}{2})}{2Mv_{\mu}r^2} \right]
{\bm \varphi}_1 ,
\label{eq:bdg3}
\end{eqnarray}
and ${\bm \varphi}_2 (r) \!\propto\! {\bm \varphi}^{\ast}_1 (r)$. Here, we set $v_{\mu}\!\equiv\!v_{\mu}(\alpha)$ and ${\bm \varphi}_1\!\equiv\! {\bm \varphi}_1(r)$. Under the condition (i)-(iii) described above, the right hand side of Eq.~(\ref{eq:bdg3}) can be regarded as the small perturbation. We now assume the solution of Eq.~(\ref{eq:bdg3}) as
\begin{eqnarray}
{\bm \varphi}_1 (r) = {\bm \varphi}^{(0)}_1 (r) + iB_1 e^{-\chi(r)} 
\left[ \begin{array}{c} \psi _1 (r) \\ -i\psi _2 (r) \end{array} \right],
\label{eq:egnvec}
\end{eqnarray}
where ${\bm \varphi}^{(0)}_1 (r)$ is the solution when the right hand side of Eq.~(\ref{eq:bdg3}) is neglected, and $\psi _{1,2}(r)$ is the small correction to ${\bm \varphi}^{(0)}_1 (r)$, {\it i.e.}, $|\psi _{1,2}(r)|\!\ll\! 1$. Hence, one finds $\varphi^{(0)}_1 (r) = B_1 e^{-\chi(r)} [ 1, -i]^T$ with
\begin{eqnarray}
\chi(r) \equiv \frac{1}{v_{\mu}(\alpha)} \int^{r}_0 \Delta(r') dr'.
\end{eqnarray}
Since $|\psi _{1,2}(r)| \!\ll\! 1$, Eq.~(\ref{eq:egnvec}) can be also expressed as
\begin{eqnarray}
{\bm \varphi}_1 (r) \simeq B_1 e^{-\chi(r)} 
\left[ \begin{array}{c} e^{i\psi _1 (r)} \\ -ie^{i\psi _2 (r)} \end{array} \right].
\label{eq:egnvec2}
\end{eqnarray}
Within Eq.~(\ref{eq:egnvec}), one can find the solution of Eq.~(\ref{eq:bdg3}) as
\begin{eqnarray}
&& \hspace{-10mm} \psi _1 (r) = - \psi _2 (r) \equiv \psi (r) \nn \\
&& \hspace{-7mm} = - \int^{\infty}_r 
\left( \frac{E_{\bm q}}{v_{\mu}} - \frac{w(q_{\theta}-\frac{w}{2})}{2Mv_{\mu}{r'}^{2}} \right)
e^{-2\{(\chi(r') -\chi(r)\}}dr',
\label{eq:psi0}
\end{eqnarray}
where we set $v_{\mu}\!\equiv\!v_{\mu}(\alpha)$, again.

To get the solution of the BdG equation (\ref{eq:bdg2}), the wave functions in two different domains, Eq.~(\ref{eq:uv_r0}) for $r\!<\! r_c$ and (\ref{eq:hankel}) for $r\!>\! r_c$, are now matched at $r \!\sim\! r_c$. Because of the condition (i), $|q_{\theta}| \!\ll\! r_c k_F$, making use of the asymptotic forms of $J_{\nu}(z)$ and $H^{(1,2)}_{\nu}(z)$ in $z\!\gg\!|\nu|$, the wave functions for $r\!<\! r_c$ in Eq.~(\ref{eq:uv_r0}) is rewritten as
\begin{subequations}
\label{eq:solution1}
\begin{eqnarray}
\hspace{-5mm}u_{\bm q} \simeq \sqrt{\frac{2M}{\pi v_{\mu}r}}
A_u\cos{
\left( 
k_{+}r + \frac{q^2_{\theta}-\frac{1}{4}}{2k_{+}r} - \frac{2q_{\theta}+1}{4} \pi 
\right)} 
\label{eq:solution1u}
\end{eqnarray}
\begin{eqnarray}
v_{\bm q}\simeq \sqrt{\frac{2M}{\pi v_{\mu}r}}
A_v\cos{
\bigg( 
k_{-}r + \frac{(q_{\theta}-w)^2-\frac{1}{4}}{2k_{-}r}} \nn \\ 
 - \frac{2q_{\theta}-2w+1}{4} \pi 
\bigg) \hspace{10mm}
\label{eq:solution1v}
\end{eqnarray}
\end{subequations}
with $v_{\mu}\!\equiv\!v_{\mu}(\alpha)$ and $k_{\pm}\!\equiv\! k_{\pm}(\alpha)$. Also, Eq.~(\ref{eq:hankel}) with Eq.~(\ref{eq:egnvec2}) for $r \!>\! r_c \!\gg\! |q_{\theta}|/k_F$ is 
\begin{eqnarray}
\hspace{-5mm}
{\bm u}_{\bm q} \simeq 
\sqrt{\frac{2M}{\pi v_{\mu}r}} e^{-\chi(r)}
\left[
\begin{array}{c}
\displaystyle{B_1e^{i\eta _{+}(r)} + B_2 e^{-i\eta _+(r)}} \\
\displaystyle{-iB_1e^{i\eta _{-}(r)} + i B_2 e^{-i\eta _-(r)}}
\end{array}
\right]
\label{eq:solution2}
\end{eqnarray}
with
\begin{eqnarray}
\eta _{\pm}(r) \equiv Mv_{\mu}r + \frac{m^2 - \frac{1}{4}}{2Mv_{\mu}r} 
- \frac{2m+1}{4}\pi \pm \psi (r) .
\end{eqnarray}

To match two expressions of $u_{\bm q}(r)$ in Eqs.~(\ref{eq:solution1u}) and (\ref{eq:solution2}) at $r\!=\!r_c$, one should put the coefficients $B_{1,2}$ as
\begin{eqnarray}
B_1 = \frac{A_u}{2} e^{i\gamma}, \hspace{3mm} B_2 = \frac{A_u}{2} e^{-i\gamma}.
\end{eqnarray}
By comparing with Eqs.~(\ref{eq:solution1u}) and (\ref{eq:solution2}), one can obtain the expression of $\psi$ as
\begin{eqnarray}
\psi(r) \simeq \frac{E_{\bm q}}{v_{\mu}}r + \frac{w(q_{\theta}-\frac{w}{2})}{2Mv_{\mu}r} 
+ \frac{m-q_{\theta}}{2}\pi - \gamma.
\label{eq:psi1}
\end{eqnarray}
In a same way, one finds the another expression from Eqs.~(\ref{eq:solution1v}) and (\ref{eq:solution2}) 
\begin{eqnarray}
\psi(r) &\simeq& \frac{E_{\bm q}}{v_{\mu}}r + \frac{w(q_{\theta}-\frac{w}{2})}{2Mv_{\mu}r} 
- \frac{m-q_{\theta}+w}{2} \nn \\ 
&& + \gamma - \left( n + \frac{1}{2} \right)\pi ,
\label{eq:psi2}
\end{eqnarray}
where $n$ is the integer. The expressions on $\psi(r)$ in Eqs.~(\ref{eq:psi1}) and (\ref{eq:psi2}) becomes identical when $\gamma$ satisfies
\begin{eqnarray}
\gamma = \frac{\pi}{2} \left( m - q_{\theta} + \frac{w+1}{2} \right) + \frac{\pi}{2}n. 
\label{eq:gamma}
\end{eqnarray}

The alternative expressions of $\psi(r)$ in Eq.~(\ref{eq:psi0}) and Eq.~(\ref{eq:psi1}) with Eq.~(\ref{eq:gamma}) should be identical at $r \!=\! r_c$. Hence, we finally obtain the eigenvalue of the BdG equation (\ref{eq:bdg}), 
\begin{eqnarray}
\hspace{-5mm}
E_{\bm q} = - \left( q_{\theta}-\frac{w}{2}\right) \frac{\omega _0}{\sin(\alpha)}
+ \left( n + \frac{w-1}{2} \right) \sin(\alpha)\omega _1, 
\label{eq:energy}
\end{eqnarray}
where 
\begin{subequations}
\label{eq:analyticE}
\begin{eqnarray}
\omega _0 \equiv 
\frac{w \displaystyle{\int^{\infty}_{r_c} \frac{\Delta(r')}{k_{\mu}r'} e^{-2\chi(r')}dr'}}
     {\displaystyle{\int^{\infty}_0e^{-2\chi(r')}dr'}}, 
\end{eqnarray}
\begin{eqnarray}
\omega _1 \equiv
\frac{\pi k_{\mu}}{2M \displaystyle{\int^{\infty}_0e^{-2\chi(r')}dr'}}.
\end{eqnarray}
\end{subequations}
To estimate the order of the energy scale of $\omega _{0,1}$, let us consider the simplest case of $\Delta(r)$, that is, $\Delta(r) \!=\! \Delta _0 \tanh{(r/\xi)}$. In this situation, one find $\omega _1 \!\simeq\! \frac{\pi}{2}\Delta _0$ and $\omega _0 \!=\! g\frac{w}{2}\frac{\Delta^2_0}{E_F}$ with $\int^{\infty}_0 e^{-2\chi(r')}dr' \!=\! \xi$. Here, $g\!\sim\! \mathcal{O}(1)$. Hence, the eigenvalue $E_{\bm q}$ is composed of two different energy scales, such as $\Delta _0$ and $\Delta^2_0/E_F$. The expression in the case of a singly quantized vortex ($w\!=\! 1$) coincides to that in Ref.~\cite{cdgm}, and that with an arbitrary $w$ reproduces the results within the semiclassical approximation in Ref.~\cite{duncan}. The further details on Eq.~(\ref{eq:energy}) is described in Sec.~II B. We should mention that the integrals in $\omega _0$ and $\omega _1$ depends on the winding number of $\Delta(r)$.

\end{document}